\documentclass[twocolumn,showpacs,preprintnumbers,amsmath,amssymb]{revtex4}
\usepackage{soul}
\usepackage{dcolumn}
\usepackage{bm}
\usepackage{bbding}
\usepackage{epsfig}
\usepackage{amsmath}

\begin{document}
\title{Non-adiabatic transitions in a non-symmetric optical lattice}
\author{L. Morales-Molina and S. A. Reyes}
\affiliation{Departamento de F\'isica, Facultad de F\'isica,
  Pontificia Universidad Cat\'olica de Chile, Casilla 306, Santiago
  22, Chile}
\begin{abstract}
We study Landau-Zener interband transitions for a non-symmetric optical lattice in the presence of an external force.  We show that gain and losses of the light beam, as well as the relative occupation probabilities of the bands involved in the transitions can be accurately managed upon tuning the amplitude of the non-Hermitian component of the lattice. Exact expressions for the transition and non-transition probabilities for a non-symmetric system obtained within a two-mode approximation are provided. These equations successfully account for the main features of the transitions in the optical lattice.
We also interpret the non-conventional Bloch oscillations at criticality studied in Phys. Rev. Lett.{\bf 103}, 123601 (2009) as a series of a Landau-Zener transitions.

\end{abstract}

\pacs{73.40.Gk, 
 42.50.Xa,
 42.25.Bs, 
 11.30.Er 
 }
\maketitle
\section{Introduction}
Quantum systems described by non-Hermitian Hamiltonians with real spectra have drawn great interest in recent years \cite{Bender}. These systems are invariant under parity and time reversal symmetry (${\cal PT}$-symmetry) and their behavior is quite different from Hermitian ones.
Important issues, such as definition of an inner product, unitarity, as well as the definition of observables for this new ``quantum mechanics'' have been intensively studied \cite{Bender}. As a result, new avenues have been opened for the observation of exotic phenomena associated with non-Hermiticity.

In particular, recent developments in optics have allowed the construction of ``complex crystals'' with unique properties such as violation of Friedel's law \cite{Ober1}. These systems are realized in waveguide arrays that include gain and loss regions. Furthermore, a restricted class of complex crystals may be built so that they posses the aforementioned ${\cal PT}$-symmetry. Along these lines, very recent experiments realized for two coupled waveguides have revealed a new behavior for light propagation in ${\cal PT}$-symmetric optical systems  \cite{Chris2}. 

Additionally, driven periodic complex potentials have proven to be a fertile ground for the study of novel phenomena in wave mechanics such as non-conventional Bloch oscillations \cite{longhi}. In this paper, we consider the scenario in which a DC force is applied to a ${\cal PT}$-symmetric complex lattice. As in conventional quantum mechanics, the application of an external force may lead to interband transitions and interference phenomena. However, since in this case the time evolution does not obey the unitarity condition, such transitions are fundamentally different. Motivated by this idea, in the present work we provide a detailed analysis of the Landau-Zener transitions for a light beam propagating in a complex potential.  

For this purpose simulations of the interband transitions are performed revealing that the behavior of the relative occupation probabilities for the bands can be accurately managed upon tuning the amplitude of the non-Hermitian component of the lattice. In particular, the power can be increased or decreased during the transition process depending on the sign of the non-Hermitian part of the potential. 

Complementary to our numerical results, we develop an analytical framework that describes the transitions near the level crossings within a two-mode aproximation. Notably, the analytical results not only provide deep insight into the dynamics of the system, but also accurately reproduce the main features observed in our numerical simulations. As a special case, we interpret the non-conventional Bloch oscillations studied at criticality in \cite{longhi} as a sequence of Landau-Zener transitions.

The paper is organized as follows: In section II, we describe the model for light propagation in periodic structures with complex refractive index within the paraxial regime. Next, in section III we consider a two-mode approximation for the description of the dynamics in the complex lattice and provide analytical expressions for its non-adiabatic behavior. In section IV, we present a detailed explanation of our numerical results for the system considered in section II and compare them with the theoretical predictions obtained within the two-mode approximation. In section V we summarize our results and discuss prospects for future developments.

\section{The Model}

Within the paraxial approximation the propagation of a monochromatic light beam  with wavelength $\lambda$ through the lattice can be described by the Schr\"odinger type equation \cite{longhi2}

\begin{equation}\label{Eq:shrod}
 i\lambdabar \partial_Z \psi=-\frac{\lambdabar^2}{2n_{s}}\frac{\partial^2 \psi}{\partial x^2} + U(x)\psi-F x\equiv {\cal H}_{0}\psi-Fx \psi
\end{equation}
where $\lambdabar=\lambda/2\pi$, $n_s$ is the refractive index of the substrate. The potential is given by 
\begin{equation}\label{Eq:poten}
U(x)=n_s- n(x),
\end{equation}
where $n(x)=n(x+a)$ is the periodic refractive index profile and $a$ is the periodicity of the lattice. The time-like coordinate $Z$ is the propagation distance along the waveguides and $x$ is the corresponding transverse axis. In the last term of (\ref{Eq:shrod}), $F$ represents the gradient of the refraction index that mimics the effect of an external DC force.

For a complex potential, the imaginary part of $U(x)$ renders ${\cal H}$ into a non-Hermitian Hamiltonian. In the following we shall consider $U(x)=U_{1}\cos(2\pi x/a)+iU_{2}\sin(2\pi x/a)$, which satisfies $U(x)=U(-x)^{*}$ and thus ${\cal H}_{0}$ is ${\cal PT}$-symmetric \cite{longhi}. The spectrum of ${\cal H}_{0}$ has real values for $U_{2}<U_{2}^{crit}=U_{1}$; otherwise the bands merge giving place to pairs of complex conjugate values \cite{Chris1}. Henceforth, we consider $U_{2}<U_{2}^{crit}$, unless otherwise specified.
The eigenfunctions of ${\cal H}_{0}$ are the Floquet-Bloch band mode functions $\phi_{q,n}(x)$ with Bloch-momentum $q$ which fulfill the secular equation ${\cal H}_{0}\phi_{q,n}(x)=E_{n}(q)\phi_{q,n}(x)$, where $E_{n}(q)$ are the eigenenergies corresponding to the $n$-th Floquet-Bloch band. 
While as in real crystals $E_{n}(-q)=E_{n}(q)$, the wavefunctions under the parity transformation $x\rightarrow -x$ and $q\rightarrow -q$ obey the general relation $\phi_{q,n}^{\dagger}(x)=\phi_{-q,n}(-x)$, where $\phi_{q,n}^{\dagger}$ are eigenfunctions of ${\cal H}_{0}^\dagger$. Moreover, the orthogonality condition becomes $\int_{-\infty}^{\infty}\phi_{q,n}^{\dagger}(x)\phi_{q',m}(x) dx = d_{q,n} \delta(q-q')\delta_{n,m}$ with $d_{q,n}=\pm 1$ \cite{Chris1}.

It is important to notice that for the ${\cal PT}$-symmetric  Hamiltonian ${\cal H}_{0}$ the power of the propagating light beam remains constant provided $U_{2}<U_{2}^{crit}$. If in addition we consider the presence of an external force as in Eq. (\ref{Eq:shrod}), the $\cal{PT}$-symmetry is broken and the power is not longer a conserved quantity. Thus, under the action of the force, transitions to other bands may occur which do not conserve the power. Recently, the Gaussian wavepacket dynamics of this system has been analyzed in the presence of a small external force \cite{longhi}. In the following sections we generalize the analysis for arbitrary values of $F$. 
 
To gain understanding of the interband transition processes, we consider as input excitation a plane wave with normal incidence. Hence, the wavefunction can be written as a superposition of the Floquet-Bloch eigenfunctions, i.e. $\psi_{q}(x,Z)=\sum_{n} c_{n}^{q}(Z)\phi_{n}^{q}(x)$, where $\phi_{n}^{q}(x)=w_{n}^{q}(x)e^{iqx}$ and $w_{n}^{q}(x+a)=w_{n}^{q}(x)$. Expanding $w^{q} _{n}(x)$ in a Fourier basis, we can write
\begin{eqnarray}
\psi_{q}(x,Z) = \sum_{l}a_{l}^{q}(Z)e^{i(2kl+q)x}, ~~\text{with}~~ k=\pi/a \nonumber \\
\text{and}~~~ a_{l}^{q}(Z)=\sum_n c_n^q b_{l,n}^q.~~~~~~~~~~
\end{eqnarray}
Here, $b_{l,n}^q$ with $l=0, \pm 1, \pm 2...$ are the coefficients of the Fourier expansion for $w_{n}^{q}(x)$. Substituting this into Eqs. (\ref{Eq:shrod}-\ref{Eq:poten}) we get the dimensionless equation for the dynamics of $a_{l}^{q}$,
\begin{equation}\label{Eq:fourier}
i\frac{\partial a^{q}_{l}}{\partial z}=(2l+\tilde{q})^2a^{q}_{l}+(V_{1}+V_{2}) a^{q}_{l+1}+(V_{1}-V_{2}) a^{q}_{l-1}
\end{equation}  
where $z=Z {\cal E}_{k}/\lambdabar$, $\tilde{q}=q/ k$ and $V_{j}=U_{j}/2 {\cal E}_{k}$,  $j=1,2$ with ${\cal E}_{k}=\lambdabar^2 k^2/2n_s$.
It is convenient to write the evolution equation in the usual quantum mechanics notation
\begin{equation}\label{Eq:schrod2}
i\partial_z |a^q(z)\rangle={\bf H}|a^q(z)\rangle,
\end{equation}
where $|a^q(z)\rangle$ is the vector with components $a_{l}^{q}$ and
\begin{equation}\label{Eq:Ham}
{\bf H}=\{H_{n,l}=(2l+\tilde{q})^2\delta_{n,l}+(V_{1}+V_{2}) \delta_{n,l-1}+(V_{1}-V_{2}) \delta_{n,l+1}\}.
\end{equation} 
This Hamiltonian matrix is real but non-symmetric and has a real spectrum for values $V_{1}>V_{2}$. From this expression it is also clear that changing the sign of the amplitude of the imaginary part of the potential $V_{2}$ is equivalent to a transposition of ${\bf H}$. Note that the problem becomes analogous to the dynamics of a particle moving on a discrete chain with non-symmetric hopping amplitudes ($V_{1}+V_{2}$ and $V_{1}-V_{2}$) and on-site energies given by $(2l+\tilde{q})^2$. Clearly, for positive values of $V_{2}$, the system tends to evolve towards modes with higher energies and viceversa. This is a manifestation of the fact that there is energy gain (loss) for positive (negative) values $U_{2}$ in the optical system (\ref{Eq:shrod}).

\begin{figure}
\begin{tabular}{lc}
\includegraphics[width=7cm,height=6cm]{Fig1a.eps}\\
\includegraphics[width=7cm,height=6cm]{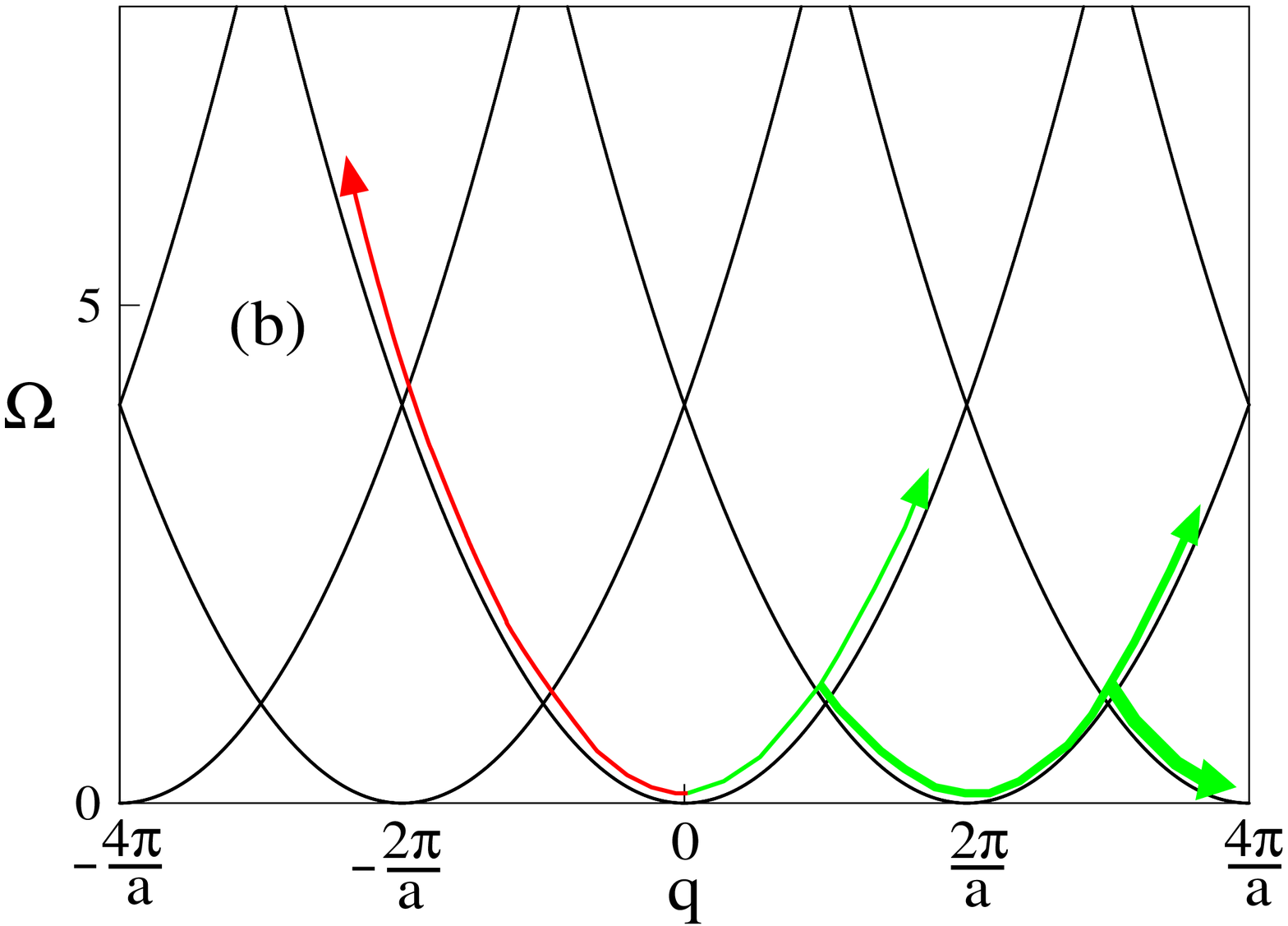}
\end{tabular}
\caption{(Color online) Energy spectrum $\Omega\equiv E/{\cal E}_{k}$ vs. Bloch-momentum $q$ for $V_{1}=0.2$. (a)  $V_{2}=0.15$; (b) $V_{2}=0.2$.} 
\label{Fig:spectrum}

\end{figure}

A numerical diagonalization of (\ref{Eq:Ham}) results in the energy spectrum shown in Fig. \ref{Fig:spectrum}. The diagram of energy vs. Bloch-momentum reveals that increasing $V_{2}$ causes an approaching of the bands at the avoided crossings leading to degeneracies at  $V_{2}=V_{1}$. 

\section{Two-mode system}

Let us focus now on the analysis near the avoided level crossings that appear at the Bragg scattering points $\tilde{q}=(2j+1)$ with $j=0,\pm1,\pm2...$. For a small energy separation between the modes involved in a given crossing, one can safely neglect the effects due to any other bands. A suitable description can then be provided in terms of an effective two-mode Hamiltonian. In particular, from  the matrix (\ref{Eq:Ham}) and considering the two lowest bands it is possible to obtain the corresponding effective Hamiltonian 
\begin{equation}\label{Eq:matrix2}
 H =
 \begin{pmatrix}
  \epsilon & \frac{\Delta+\delta}{2}  \\
  \frac{\Delta-\delta}{2}   & -\epsilon  \\
 \end{pmatrix},
\end{equation}
where $\epsilon=2(1-2l-\tilde{q})$, $\Delta=2V_{1}$ and $\delta=2V_{2}$. Unlike the Hermitian case, in Eq. (\ref{Eq:matrix2}) $H_{1,2}\neq H_{2,1}^{*}$ and the eigenvalues can therefore have complex values. By solving the secular equation for (\ref{Eq:matrix2}), we get $E_{\pm}=\pm\sqrt{\epsilon^2+\frac{(\Delta^2-\delta^2)}{4}}$ so that for high values of $\epsilon$ the spectrum is real. At $\epsilon=0$ the gap between the energy levels becomes $E_{+}-E_{-}=\sqrt{\Delta^2-\delta^2}$. On the other hand, when $\delta^2>\Delta^2$ purely imaginary eigenvalues arise in the region $\epsilon^2<(\delta^2-\Delta^2)/4$, indicating that the system is beyond the critical point where the ${\cal PT}$-symmetry is broken. 

\subsection{Landau-Zener for a non-symmetric  two-mode system}
The presence of a gradient in the refraction index in Eq. (\ref{Eq:shrod}) mimicking a DC force causes a linear variation of the Bloch-momentum along the propagation distance, i.e. $\tilde{q}=\alpha z$, where $\alpha$ is proportional to the force. As a result the diagonal elements of matrix (\ref{Eq:matrix2}) are modified as $\epsilon=-2\alpha z+\epsilon_{0}$, where $\epsilon_0=2(1-2l)$. By introducing the change of variable $t= z-\epsilon_0/2\alpha$  we obtain $\epsilon= -\frac{\beta t}{2}$ where $\beta=4\alpha$. Thus, at $t=0$ the energy levels are only separated by the gap. In Landau-Zener theory the variable $t$ represents time and $\beta$ is the speed for the variation of $\epsilon$ whose tuning controls the transitions between the two bands \cite{Zener,Sevchenko}. To study the interband transition, the initial state is prepared away from the avoided crossing where $|\beta t|\gg \sqrt{\Delta^2-\delta^2}$. In the conventional Landau-Zener framework this amounts to a preparation at $t\rightarrow -\infty$. For simplicity we consider $|a_{1}(-\infty)|^2=1$ for the lowest energy state and  $a_{2}(-\infty)=0$ for the excited state.

Now, as the system evolves, we find that after sufficiently long time the asymptotic values for the intensities are given by \cite{Seba} 
\begin{equation}\label{Eq:Zener2}
|a_{2}(\infty)|^2=P=\exp\left[{-\pi \frac{\Delta^2-\delta^2}{2\beta}}\right]
\end{equation}
for the excited state and
\begin{equation}\label{Eq:Zener1}
|a_{1}(\infty)|^2=\Gamma(1-P )
\end{equation}
for the ground state,
where $\Gamma=(\Delta+\delta)/(\Delta-\delta)$. 

Since for this two mode system the power is given by $\rho=|a_{1}|^2+|a_{2}|^2$, after the transition we have that $\rho=\Gamma(1-P)+P$, which in the limit of very slow variation $\beta \ll (\Delta^2-\delta^2)$ reduces to $\rho=|a_{1}|^{2}=\Gamma$. 
Thus, permutation of the nondiagonal matrix elements in (\ref{Eq:matrix2}) results in power variations of the order of $4\Delta\delta/(\Delta^2-\delta^2)$. This shows that even a slow navigation through the spectrum can greatly affect the power due to the coupling with the upper state. In this regard, the quantities $|a_2|^2 $ and $|a_{1}|^2$ can only be referred to as the relative transition and non-transition probabilities, respectively.

To get some insight into Eq. (\ref{Eq:Zener1}), we analyze several regimes for the parameter $\delta$. First of all, if we set $\delta=0$ into the equations above, corresponding to a real potential ($V_{2}=0$), the original Landau-Zener probability transition $P_{LZ}=\exp[-\pi \Delta^2/2\beta]$ is recovered and the non-transition probability becomes $(1-P_{LZ})$. If $0<\delta<\Delta$, the relative transition probability $P$ increases for increasing values of $\delta$. The same holds for the quantity $\Gamma(1-P)$, even though the quantity $(1-P)$ is decreased. On the other hand, for $-|\Delta|<\delta<0$ and increasing $|\delta|$, the relative nontransition probability decreases.
\begin{figure}
\includegraphics[width=9cm,height=6cm]{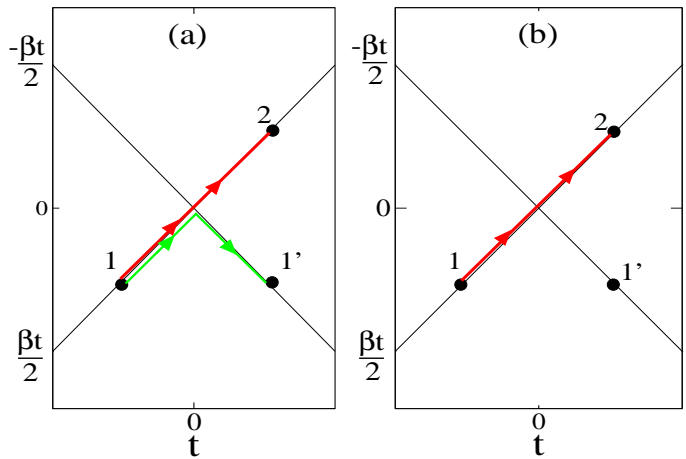}
\caption{(Color online) Schematic representation of two energy levels in the limit $|\delta|=\Delta$. (a) $\delta=\Delta$; (b) $\delta=-\Delta$.} 
\label{Fig:crossing}
\end{figure}

Consider now the limiting case $\delta \rightarrow \Delta$ with $\beta\neq 0$ where the gap vanishes, i.e. there is a band crossing. In this regime the expressions given in (\ref{Eq:Zener1}-\ref{Eq:Zener2}) reduce to 
\begin{equation}\label{Eq:limit1}
\lim_{\delta \rightarrow \Delta}|a_{1}|^2=2 \pi \frac{\Delta^2}{\beta}\equiv\Lambda, \,\,\,\,\,\ |a_{2}|^2=1.
\end{equation}
Observe that for small values of $\beta$ the intensity in the lowest mode is greatly increased for any $\Delta>0$. In contrast, for negative $\delta$, the asymptotic intensities become
\begin{equation}\label{Eq:limit2}
\lim_{\delta \rightarrow -\Delta}|a_{1}|^2=0, \,\,\,\,\,\ |a_{2}|^2=1.
\end{equation}

It is interesting to note that, contrary to the conventional Landau-Zener transition  process for a vanishing gap, there is a non-zero occupation probability for the ground state, as observed in the first scenario (\ref{Eq:limit1}). On the other hand, the relative transition probability to the excited state is $|a_{2}|^2=1$ (see Fig. (\ref{Fig:crossing}a)). Remarkably, in the second scenario, there is a complete transition to the excited state with $|a_{1}|^2=0$ (see Fig. (\ref{Fig:crossing}b)) as expected for a standard Landau-Zener process. The fundamental differences between these two situations are a manifestation of the non-symmetric form of the system (\ref{Eq:matrix2}).

So far we have shown the results for a non-adiabatic transition for a single two-level system. Based on the same scheme, these results can be applied to the description of a series of avoided crossings in the optical multilevel system (\ref{Eq:Ham}). Let us now assume that the input excitation lies completely in the lowest mode, i.e. $|a_{j}|^2=\delta_{1j}$. As the force $F$ is turned on, the bloch momentum is modified and the system undergoes a succesion of transitions through several (avoided) level crossings. Note that due to the lack of coupling among higher bands, the only allowed transitions are those between the two lowest modes. Consequently, it is straightforward to find that for $F\neq0$ the total power after $n$-th consecutive avoided crossings is given by
\begin{equation}\label{Eq:norma}
 \rho_{n}=\Gamma^{n}(1-P)^{n} + P\sum_{i=0}^{n-1} \Gamma^{i}(1-P)^{i},
\end{equation}
whose expression in the limit $\delta \rightarrow \Delta$ reduces to $\rho_n=\sum_{i=0}^{n}\Lambda^i$ (cf. Eq. (\ref{Eq:limit1})). Note that the power always exceeds unity. On the other hand, taking $\delta \rightarrow -\Delta$ the power reduces to $\rho_n=1$. 

In expression (\ref{Eq:norma}) the first term accounts for the change of probability density in the lowest level, while the second term represents the part of the  probability density that is transferred to higher energy levels after $n$ avoided crossings.  

 \begin{figure}
\begin{tabular}{lc}
\includegraphics[width=7cm,height=6cm]{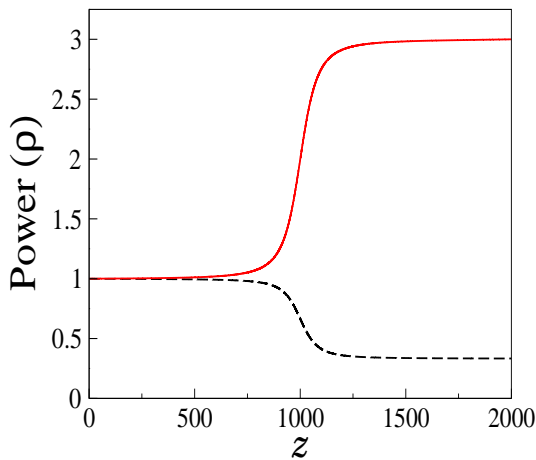}
\end{tabular}

\caption{(Color online)
Power ($\rho$) as a function of $z$.  $|\alpha|=1\times 10^{-3}$ and $V_{2}=0.1$. solid line:$\alpha>0$, dashed line: $\alpha<0$.} 
\label{Fig:fit} 
\end{figure}

\section{$\cal{PT}$-symmetric optical waveguide arrays}
The optical realization of a medium with ${\cal PT}$-symmetric properties has paved the way for the synthesis of artifical optical periodic lattices with ${\cal PT}$-symmetry. This makes the observation of interband transitions in a ${\cal PT}$-symmetric lattice a target within reach in a near future.  
 
As explained above interband transitions are generated by the presence of a gradient in the refractive index. 
In order to understand these proceeses we look at the transition points, i.e. the avoided energy crossings of the band structure of the complex lattice. 
Beam splitting occurs at these scattering points due to non-adiabatic transitions and the intensity of the beam is modified in the process. We focus our analysis on the parameter window $0 < V_{2} < V_{2}^{crit}$, where there are no singular or exceptional points. As previously noted, the action of the fictitious force in Eq. (\ref{Eq:shrod}) can be effectively replaced in Eq. (\ref{Eq:fourier}) by a Bloch-momentum $\tilde{q}$  that depends linearly on $z$, i.e. $\tilde{q}=\alpha z$ with $\alpha= F/( k{\cal E}_k)$. 

As the initial state we choose an input excitation that populates the lowest band with Bloch-momentum $\tilde{q}=\tilde{q}_{ini}=0$, which can be realized by shining a normal incident plane wave in the $z$ direction. We start considering first the limit of slow variation, $|\alpha|\ll (V_1^2-V_2^2)$.
Fig. (\ref{Fig:fit}b) shows the numerical results for the evolution of the power $\rho$ in this limit for both positive and negative $\alpha$.
Interestingly, for $\alpha>0$ the asymptotic value for the power is $\rho^{+}\approx3$ whereas for $\alpha<0$ is $\rho^{-}\approx 1/3$. Moreover, these values can also be accurately estimated using Eq. (\ref{Eq:norma}) in the corresponding limit for a single avoided crossing ($n=1$), giving $\rho^{+}=3$ and $\rho^{-}=1/3$ respectively.

\begin{figure}
\includegraphics[width=7cm,height=6.5cm]{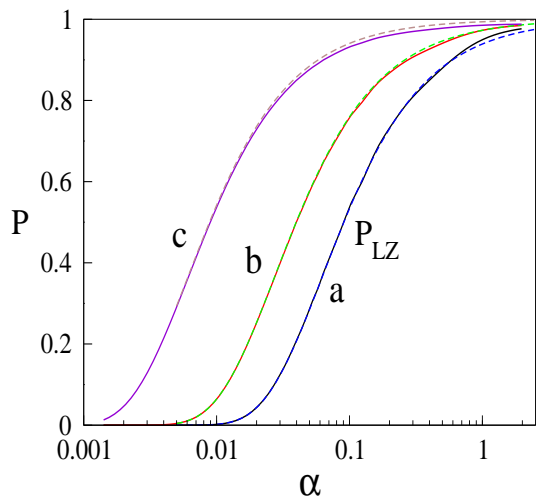}
\caption{(Color online) Probability transition $P$ vs. rate $\alpha$. (a) $V_{2}=0$, (b) $V_{2}=0.15$, (c) $V_{2}=0.19$. Dashed lines superimposed correspond to the results obtained with Eq. (\ref{Eq:Zener2}).} 
\label{Fig:Prob}
\end{figure}   
Next, for a complete analysis of the interband transition process, we consider a range of values of $\alpha$ going from the adiabatic limit considered before, to a strongly non-adiabatic regime where the energy associated to the force becomes much larger than the interband gap. To do so, for each $\alpha$ we perform simulations for the evolution of the system from $\tilde{q}_{ini}=0$ to $\tilde{q}_{fin}=1.8$ and measure the resulting occupation of the excited state. The latter is obtained by projecting the evolved wavefunction $\Phi$ onto the eigenstate $\phi_{q,2}^{\dagger}$ corresponding to the second lowest band of $\cal{H}^{\dagger}$ with $\tilde{q}=\tilde{q}_{fin}$ leading to $P=|\langle\phi_{q,2}^{\dagger} |\Phi\rangle|^2$.
In Fig.\ref{Fig:Prob} we plot the numerical calculations for the transition probability $P$ vs. $\alpha$. It is observed that the profile for $P(\alpha)$ in the non-symmetric case remains very similar to the one for $V_{2}=0$ if the system is far from the critical point ($V_{2}\ll V_2^{crit}$). In fact, the difference between the non-symmetric and symmetric cases becomes noticeable only when $V_2$ is fairly close to its critical value. This happens because as $V_{2}$ approaches $V_{2}^{crit}$, the gap is drastically reduced, leading to a strong increase in the transition probability. As the limit $V_{2}=V_{1}$ is reached, the gap vanishes and the transition probability becomes unity for any value of $\alpha\neq0$. Using Eq. (\ref{Eq:Zener2}) derived within the two-mode model, we obtain the corresponding results which are also plotted in Fig. (\ref{Fig:Prob}).
Notably, the approximated analytical results match the numerical simulations impressively well, complementing the agreement found in the adiabatic limit. 

\begin{figure}
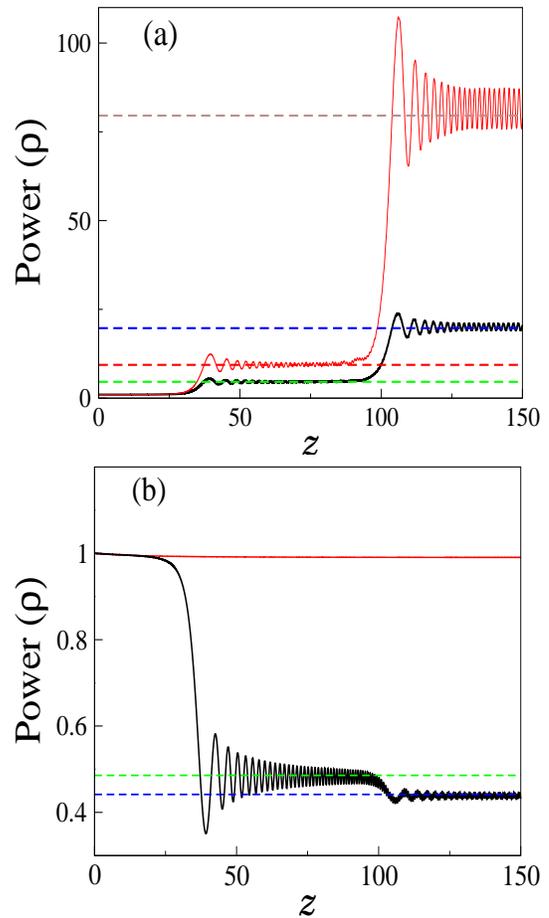

\begin{tabular}{lc}
\includegraphics[width=7cm,height=6cm]{Fig5a.eps}\\
\includegraphics[width=7cm,height=6cm]{Fig5b.eps}
\end{tabular}
\caption{(Color online) Power $\rho$ as a function of the propagation distance $z$. (a) $\alpha>0$;   (b) $\alpha<0$;   In both panels solid lines represent: Critical case $V_{2}=V_{1}$ (thin red line) and $V_{2}=0.15$ (thick black line). Dashed lines mark the asymptotic  values calculated with the corresponding Eq. (\ref{Eq:norma}). The other parameters are $V_{1}=0.2$, $|\alpha|=3\times 10^{-2}$ and the critical case is computed with $V_{2}=0.2-10^{-7}$. } 
\label{Fig:norma}
\end{figure} 

As the system is driven through Bloch-momentum space, a series of avoided crossings in the spectrum may lead to multiple transitions. In this regard, we carried out simulations going from $\tilde{q}_{ini}=0$ to $\tilde{q}_{fin}=3.9$ with $\alpha=3\times 10^{-2}$, such that the window in the $q$ axis includes two avoided crossings. Fig. \ref{Fig:norma} shows the power $\rho$ as a function of the propagation distance $z$. The fundamental feature in both panels is the emergence of steps at the Bragg scattering points as the power evolves. While for $\alpha>0$ (Fig. \ref{Fig:norma}a) the power exhibits growing steps after every avoided crossing, in the case of $\alpha<0$ the power decreases in a sequence of small steps (Fig. \ref{Fig:norma}b). To understand these two different scenarios we follow the evolution of the beam power through the band struture with the help of expression (\ref{Eq:norma}). 

Let us first analyze the critical situation $V_{2}=V_{1}$ where the two lowest bands cross: For a positive force our analytical results predict a behavior similar to Bloch oscillations, reflected in the large value of the intensity of the ground state after $n$ consecutive crossings ($\Lambda^{n}$) and with the remaining power ($\sum_{i=0}^{n-1}\Lambda^i$) transferred to higher energy bands \cite{longhi}. This process is schematically depicted in Fig. \ref{Fig:spectrum}b by arrows pointing to the right. In contrast, for a negative force, the ground state has a null intensity after $n$ succesive crossings which clearly indicates the absence of Bloch oscillations. Here, a full transition to the excited state takes place after undergoing the first crossing and since there is zero coupling between higher bands, the state follows a parabolic trajectory as schematically depicted in Fig. \ref{Fig:spectrum}b with arrows pointing to the left.

For nonzero gap, transitions become determined by the tradeoff between the applied force and the gap, resulting in a rich picture of Bloch-oscillation-like behavior combined with interband transitions whose average power can be estimated using expression (\ref{Eq:norma}) (see horizontal dashed lines in Fig. \ref{Fig:norma}). Notice that for $\alpha>0$ the height of the steps is decreased when the amplitude of the imaginary component of the potential is reduced (Fig. \ref{Fig:norma}a), as expected due to the associated absorption process. Conversely, for $\alpha<0$ there is an attenuation of the power at each transition as observed in Fig. \ref{Fig:norma}b. Once again the applicability of the theoretical equations derived within the two-mode formalism is demonstrated by the excellent agreement with numerical simulations. 

We would also like to point out that all of the results mentioned above for the two-mode approximation correspond to asymptotic values far from the avoided level crossing. Nevertheless, close to the transition region, the exact two-mode calculation exhibits an oscillating behavior similar to the one present in the simulations shown in Fig. \ref{Fig:norma}. Clearly, these oscillations arise as a result of the interaction between the bands and are damped out as the coupling becomes negligible when the system evolves away from the (avoided) level crossing.

\section{Concluding remarks}

In summary we have investigated the Landau-Zener transitions for a $\cal{PT}$-symmetric complex lattice that can be realized in optical waveguide arrays. Firstly, the analysis was performed using an exact calculation for the two-mode approximation and later extended to the complete multi-mode system driven through a series of avoided crossings. Overall, we found that the two-mode framework provides a simple and transparent understanding of the main features of the dynamics in the complex crystal.

The analytical expressions presented above not only reflect the non-reciprocity of the lattice, but also accurately predict the simulation results including the more general  scenario with multiple crossings. Also, it is found that the power is very susceptible to the system speed as it moves accross the Bragg scattering points, where the passage time determines the amount of energy absorbed/emitted by the light beam. This mechanism can be used as a way to control the intensity of the light beam in a complex waveguide array.

In a different context, complex optical lattices for cold atoms have been built \cite{Ober}. In this regard, investigation of transitions for matter waves between bands realized for real optical lattices \cite{arimondo,Weitz1,Weitz2} could in principle be extended to ${\cal PT}$-symmetric systems.
Furthermore, the results exposed on this work could eventually serve as a benchmark for future developments in the several areas that involve Landau-Zener processes and interferometry for non-symmetric Hamiltonians.

\acknowledgements{
The authors acknowledge financial support from  Fondecyt  project No. 1110671. 
LMM is also supported by the start-up funding proyecto Inicio.}

\end{document}